\def\deg      {{\ifmmode^\circ\else$^\circ$\fi} } 
\def\arcm     {{\ifmmode {'\ }\else$'     $\fi} } 
\def\arcs     {{\ifmmode{''\ }\else$''    $\fi} } 

\def\Msun     {{\ $M_{\odot}$} }

\documentclass[namedreferences]{kluwer}

\usepackage{epsfig}

\begin{document}
\begin{article}
\begin{opening}

\title{Galaxy Evolution in the Reddest Possible Filter}

\author{E. A. \surname{Richards}\email{erichard@nrao.edu}}
\institute{National Radio Astronomy Observatory \&
University of Virginia, 520 Edgemont Road, Charlottesville, 
VA 22903, USA}

\begin{abstract}

        We describe an observational program aimed at
understanding the radio emission from distant, rapidly
evolving galaxy populations. These observations were
carried out at 1.4 and 8.5 GHz with the VLA centered on
the Hubble Deep Field,
obtaining limiting
flux densities of 40 and 8 $\mu$Jy respectively.
The differential count of the radio sources is
marginally sub-Euclidean to the completeness limits
($\gamma$ = -2.4 $\pm$ 0.1) and fluctuation
analysis suggests nearly 60 sources
per arcmin$^2$ at the 1 $\mu$Jy level. Using
high resolution 1.4 GHz observations obtained
with MERLIN, we resolve all radio sources
detected in the VLA complete sample and
measure a median angular size for the
microjansky radio population of 1-2\arcs .
This clue coupled with the
steep spectral index of the 1.4 GHz selected
sample suggests diffuse synchrotron
radiation in $z\sim$1 galactic disks.

        The wide-field HST and ground-based
optical exposures show that the radio
sources are identified primarily with
disk systems composed of irregulars, peculiars,
interacting/merging galaxies, and a few isolated
field spirals. Only 20\% of the radio sources
can be attributed to AGN -- the majority are likely
associated with starburst activity. The available
redshifts range from 0.1-3, with a mean of
about 0.8. We are likely witnessing a major
episode of starburst activity in these
luminous (L $>$ L$*$) systems, occasionally
accompanied by an embedded AGN.

	About 20\% of the radio sources remain
unidentified to I = 26-28 in the HDF and
flanking fields. Several of these objects
have extremely red counterparts. We suggest
that these are high redshift dusty protogalaxies.

\end{abstract}
\end{opening}

\section{A Radio Perspective of Galaxies}

        Determining how galaxies form and
subsequently evolve remains a subject of intense
study despite decades of research.
Traditionally, astronomers primarily have used
optical methods to study the characteristics
of local and distant galaxies. 
With the discovery of
the infra-red ultraluminous galaxies in the
early 1980s (e.g., Soifer et al. 1984), it was soon recognized
that optical studies can severely bias the
understanding of galaxy properties due to
dust obscuration, especially in
those systems undergoing enhanced episodes
of star-formation activity (i.e., a starburst galaxy).

        The far infrared (FIR) radiation
(30 $\mu$m $< \lambda <$ 300 $\mu$m) of
a starburst is composed of reprocessed
ultraviolet(UV) and optical light
from young, recently formed  stellar
populations. This radiation is absorbed
by dust in the interstellar medium
and thermally reradiated at FIR
wavelengths.
Closely related to the FIR emission
in starburst galaxies is the radio continuum.
Although the radio emission
is linked to active star-formation by
different physical mechanisms than that
of the FIR, there is a tight correlation
between the FIR and radio luminosity of
a starburst (Helou et al. 1985).

    In normal galaxies (i.e., without a powerful AGN),
the centimeter radio luminosity is dominated
by diffuse synchrotron emission believed to be
produced by relativistic electrons accelerated
in supernovae remnants.
Detailed radio studies
of nearby starburst galaxies such as M82 (Kronberg et al. 1985,
Muxlow et al. 1994) and
Arp 220 (Smith et al. 1998)
have revealed large numbers of young radio supernovae, embedded in
extended synchrotron haloes formed by a combination of old, coalesced
SNRs and cosmic ray injection into the surrounding disks of these
galaxies. 
As the  synchrotron radiation of a
starburst dissipates on a
physical time scale of $10^7-10^8$ years, the
radio luminosity is a true measure of the
{\em instantaneous} SFR in a galaxy, uncontaminated by
older stellar populations. Furthermore, since supernovae
progenitors are dominated by $\sim$8 \Msun stars,
synchrotron radiation has the additional advantage of being
less sensitive to uncertainties in the initial
mass function as opposed to UV and optical
recombination line studies. 
Because galaxies and the inter-galactic
medium are transparent at centimeter wavelengths,
radio emission is a sensitive measure of
star-formation in distant galaxies.

\begin{figure}
\centerline{\epsfig{file=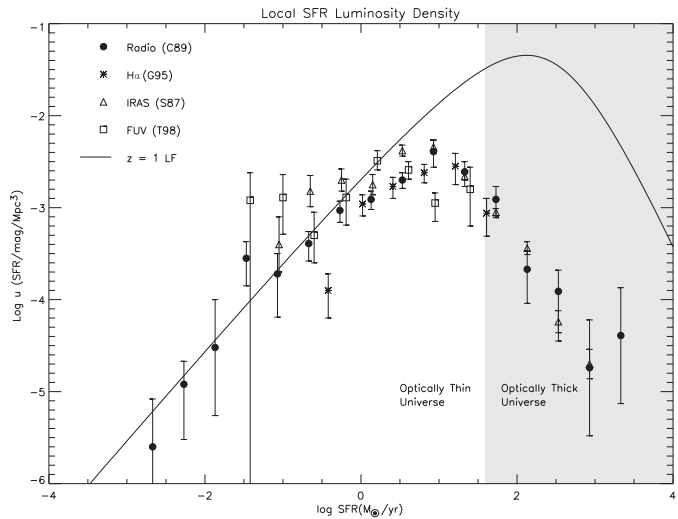,width=25pc}}
\caption{Shown is the contribution to the local
star-formation luminosity density (u) per luminosity interval of
star-forming galaxy. The radio, IRAS, H$\alpha$, and
far-ultraviolet (FUV) luminosity functions 
have been converted to SFRs
assuming a Salpeter IMF over 0.1-100 \Msun .
The shaded region represents
what may be a dust curtain beyond which optical
surveys are blind to star formation.
If the SFR luminosity function
evolves as L $\propto (1+z)^{3.5}$, then by $z$ = 1,
it will appear as the
solid line. This analysis suggests that the
bulk of global star-formation at high$-z$ is hidden from optical
surveys.}
\end{figure}

Comparison of the local radio
luminosity function (LF) of star-forming galaxies
(Condon 1989)
with those derived independently at
FIR (Soifer et al. 1987), H$\alpha$ (Gallego
et al. 1995), and UV  wavelengths (Treyer et al. 1998)
shows good agreement (Cram 1998).
Figure 1 shows the four
LFs in units of SFRs. This analysis
suggests the bulk
of local star formation is occurring in
modest starbursts with SFR $\sim$ 10 \Msun yr$^{-1}$.
However, past the peak in the
LF, the H$\alpha$ and UV estimates
begin to fall below the radio/FIR rates,
and at about 50\Msun yr$^{-1}$ has
entirely vanished.
The radio source counts and redshift statistics
are both consistent
with pure luminosity evolution of the local
population to $z \sim$ 1 with L $\propto (1+z)^{3.5}$
(e.g., Rowan-Robinson et al. 1993).
Thus the peak
in the star-forming
RLF at $z \sim$ 1 likely moves past a few hundred
 \Msun yr$^{-1}$ where optically selected surveys
become severely biased. Deep radio surveys, sensitive
to star-forming galaxies to $z \sim$2, provide unique
information on distant, rapidly evolving galaxy populations.

\section{Radio Observations of the Hubble Deep Field}

        The HDF has been observed previously
with the VLA at 8.5 GHz to a one sigma sensitivity
of 1.8 $\mu$Jy (Richards et al. 1998).
In June 1997 we observed the HDF region for
an additional 40 hours at 8.5 GHz.
We mosaiced an area defined by four separate pointings
offset from the center of the HDF by the half-power
point of the primary beam response (2.7\arcm
)
in the cardinal directions for about 10 hours
duration each. 
The final combined 8.5 GHz images have
an effective resolution of 3.5\arcs ~and
a completeness limit of 8 $\mu$Jy which
rises to 40 $\mu$Jy at 6.6\arcm from the
HDF center. Within this area we detected
40 sources in a complete (5 $\sigma$) 
with an additional 19 sources in a supplementary
sample (3.5-5 $\sigma$). 

	In November 1996, we observed the HDF
at 1.4 GHz with the VLA. The observational
details and data processing are discussed
by Richards (1999). The 1.4 GHz VLA image 
covers 40\arcm ~diameter with
an effective resolution of 1.8\arcs ~
and an rms noise of 7.5$\mu$Jy. We defined
a completeness limit of 40$\mu$Jy to
compose a catalog of 371 radio sources of
which 30 were detected at 8.5 GHz. In total
16 radio sources lie in the HDF.

	In February 1996 and April 1997, we
observed the HDF with the MERLIN interferometer
at 1.4 GHz for a total of 17 $\times$ 24 hours. These
data were combined with the VLA data to
produce sky images around all 89 previously
known radio
sources in the inner 10\arcm $\times$ 10\arcm 
of the field. These high resolution 1.4 GHz
images have a rms noise of 3.3 $\mu$Jy at
0.2\arcs ~resolution (Muxlow et al. 1999). 

\section{Radio Angular Sizes and Spectra}

	All previous high resolution studies
of submillijanksy radio sources have been
limited to approximately 2\arcs resolution.
Because the median angular size is known to
change rather sharply below a few millijansky
at 1.4 GHz (presumably due to the emergence of
a new population) from $\sim$10\arcs ~to
a few arcsec, our present observations are
uniquely suited to study the radio morphologies
of the faintest radio sources for the first time.

\begin{figure}
\centerline{\epsfig{file=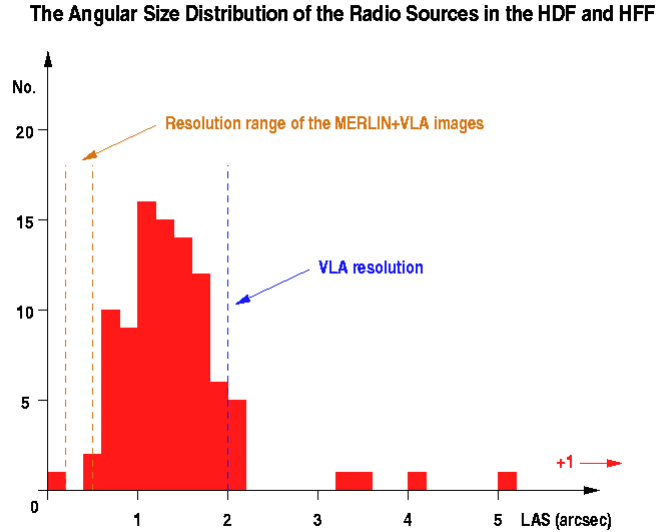,width=25pc}}
\caption{Angular sizes of sources detected
in the VLA+MERLIN images.}

\end{figure}

	Figure 2 shows the angular sizes of 
the 89 sources detected by both MERLIN and
the VLA. Virtually all radio sources are 
resolved at 0.2\arcs resolution.
There are very few radio sources with 
sizes greater than a few arcsec which
are generally associated with classical
FR I and II radio galaxies. Rather, 
the median angular size for our sample
is between 1-2\arcs , indicative of
radio emission on galactic or sub-galactic
scales. 

	The spectral index (S $\propto$ $\nu ^{-\alpha}$)
of a source can be used to diagnose the origin
of the radio emission. Inverted spectrum sources
are invariably associated with self-absorbed synchrotron 
emission associated with an AGN.
Flat spectrum sources (0 $< \alpha$ 0.5) can be
produced by AGN or optically thin Bremsstrahlung
radiation from star-formation at higher ($\nu >$ 5 GHz)
radio frequencies. Steep spectrum sources ($ \alpha >$ 0.5)
consist of diffuse synchrotron emission, often associated
either with radio jets or star-formation in galaxies.

	For the 8.5 GHz selected sample in the
HDF, the median spectral index is ${\alpha } _{8.5}$ = 
0.35 . Less than 15\% of these sources are
inverted. Although several of these
radio sources are dominated by an AGN, many
show diffuse radio emission which is likely
a combination of diffuse synchrotron and
free-free radiation associated with wide-scale
star formation. The 1.4 GHz selected sample
has a median spectral index ${\alpha } _{1.4}$ =
0.85 . Thus the microjanksy radio population
at 1.4 GHz is dominated by sources with
diffuse synchrotron emission.

\section{Optical Identifications}
	
	The absolute astrometric accuracy
of our interferometric images is set by 
our phase calibrator which has a position
error of 0.02\arcs . The independent VLA and MERLIN 
radio positions for sources detected in the
HDF agree to 0.04\arcs . The HDF WFPC2 
images were previously aligned to the radio
reference frame to about 0.1\arcs ~accuracy
(Williams et al. 1996). We bootstrapped
each of the eight individual WFPC2 
flanking field images to the radio grid
by first aligning a widefield I-band image
provided to us by Barger et al. (1999)
with our radio sources.
Optically bright galaxies detected in
both the HST and the ground-based images
were used to register the individual 
WFPC2 frames to an accuracy of 0.1-0.2\arcs .

\begin{figure}
\centerline{\epsfig{file=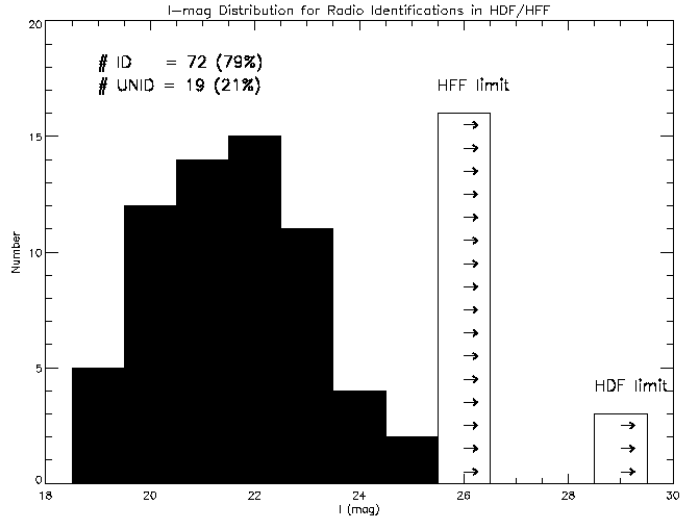,width=26pc}}
\caption{I magnitude histogram of radio sources in
the HDF region.}
\end{figure}

	Of the 91 radio sources contained in
published optical images (Barger et al. 1999),
72 have clear identifications with reliabilities
ranging from 95-99\% (Richards et al. 1998). Figure
3 presents the magnitude histogram of these 
galaxies. The mean of the identifications is
I = 22 mag, with a clear decline in the
distribution past I = 23 mag. Sixteen radio
sources cannot be identified in the HFFs or
ground-based images, and three fields are blank
in the HDF itself.

	Of the radio sources identified, the majority
reside in disk systems composed of mergers, irregulars,
and/or isolated spiral galaxies (Richards et al. 1998).
A few red ellipticals
are apparent which are almost certainly
AGN. Of the radio sources with redshifts, most are
at $z$ = 0.4-1 but we caution that there still exists
70\% incompleteness in the sample. Many of the 
disk systems identified as radio sources have clear
indications of active star-formation, including
prominent but narrow emission lines ([OII] and H$\alpha$),
mid-infrared excesses (Aussel et al. 1999) or 
peculiar optical morphology. These clues coupled
with the diffuse, steep spectrum radio emission
give strong evidence that the {\em majority of
radio sources in the HDF region are starburst
galaxies.} The implied starformation rates for
those galaxies with redshifts range from
10\Msun /yr to 1000s \Msun /yr. Two of the radio
starbursts we typically detct are shown in Figure 4. 

\begin{figure}
\centerline{
\epsfig{file=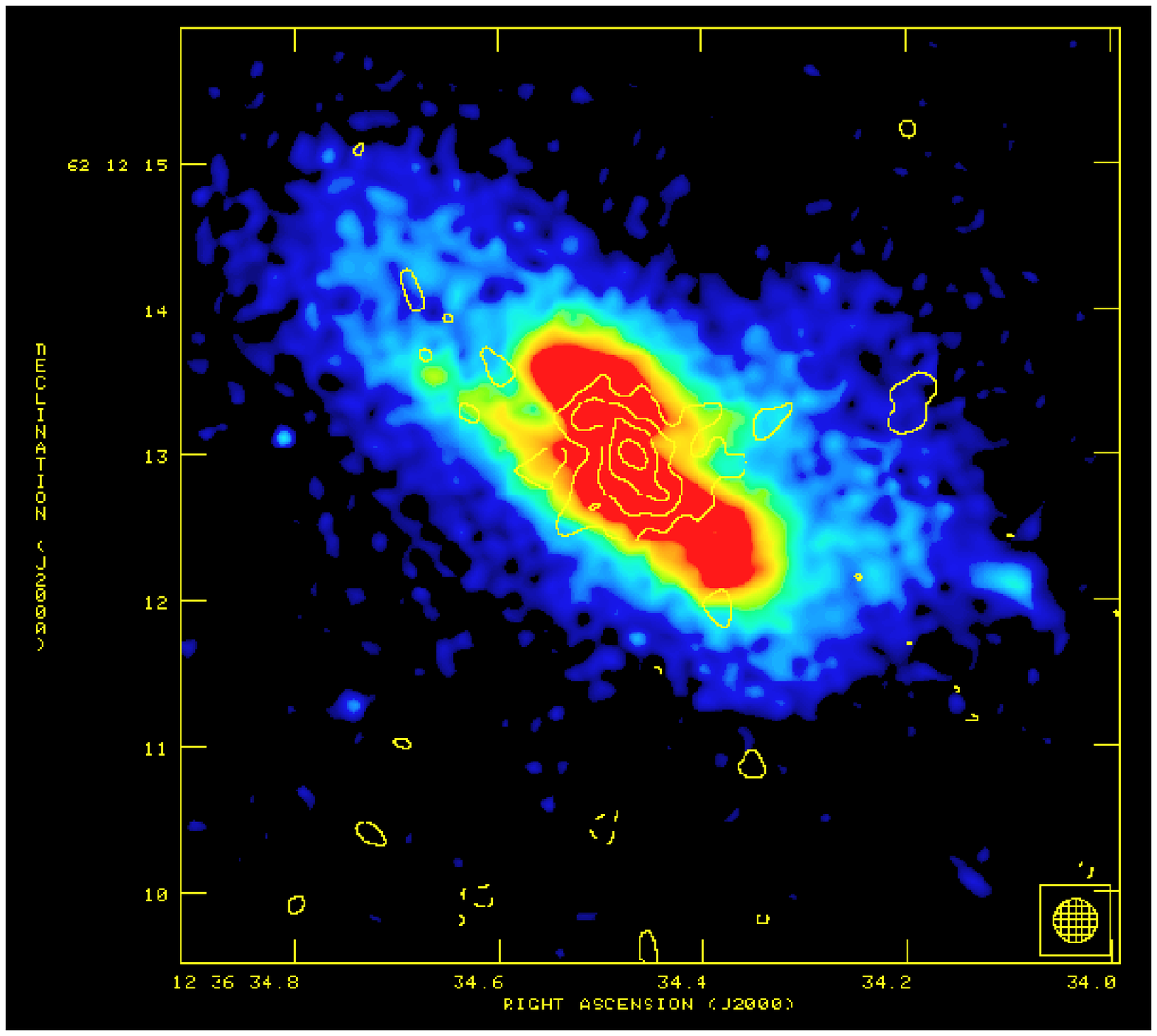,width=20pc}\qquad
\epsfig{file=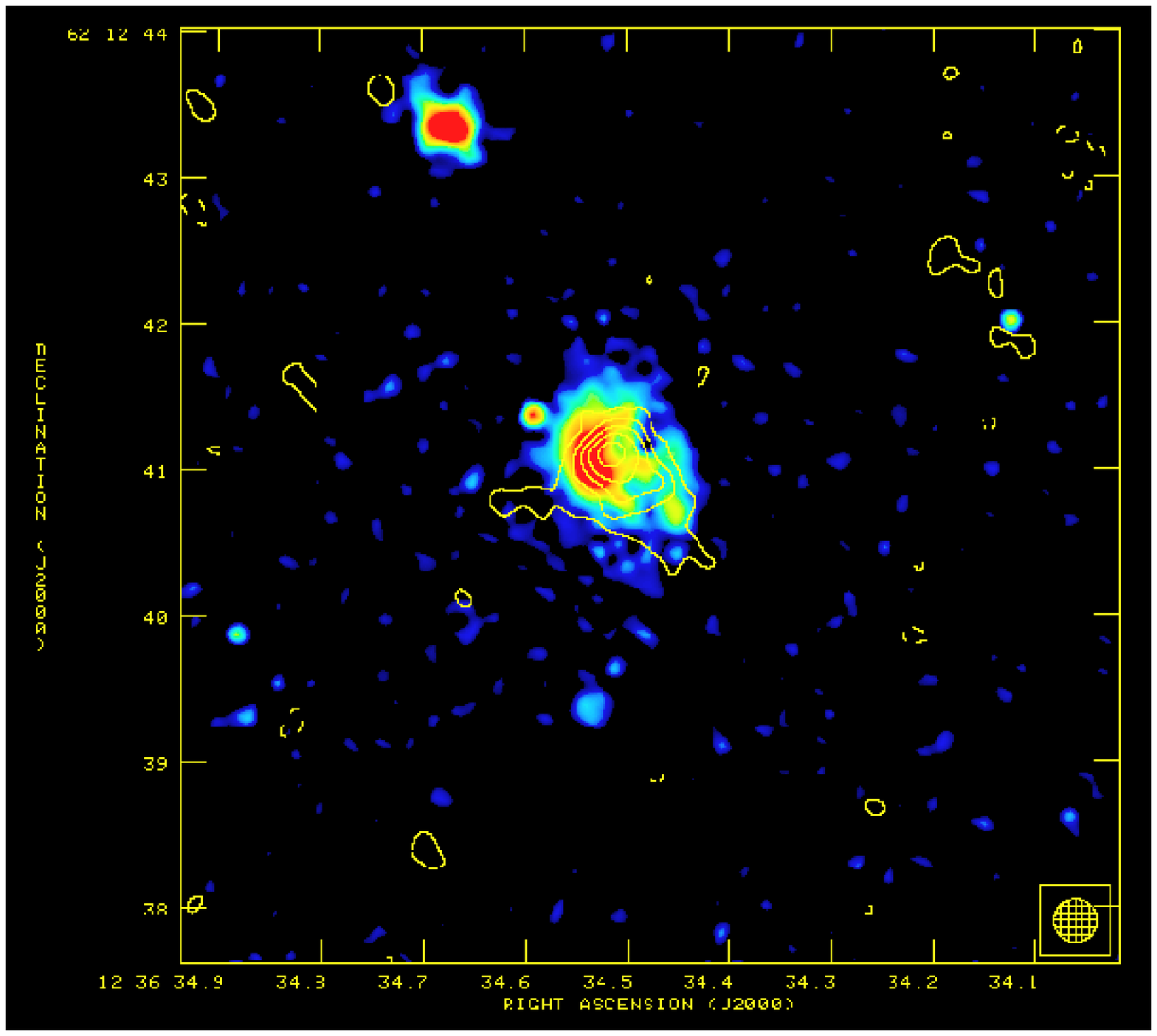,width=20pc}}
\caption{Left) Radio contours for VLA J123634+621212 are
overlaid on the WFPC2 image of this I = 19 mag. 
merging system at $z $ = 0.46 (Cohen et al. 1996). This radio source 
also has a firm ISO detection (Aussel et al. 1999). The radio SFR 
= 140 \Msun /yr. The radio emission peak is coincident
with a prominent dust lane. Right) VLA J123634+621240 is
another candidate merger at $z $ = 1.22 (Cohen 96)
and with a bright ISO detection (Aussel et al. 1999).
The steep spectrum radio emission implies a SFR = 2100 \Msun /yr .
}
\end{figure}

\subsection{Candidate High-$z$ Radio Sources}

	Although the majority of radio sources in the
HDF region can be identified with rather bright
optical galaxies (I $\sim$ 22 mag.), 20\% of
the sources remain in blank fields. These sources range
in significance from 6 - 100 $\sigma$. Deep infrared imaging
exists for several of these radio sources 
(Barger et al. 1999, Waddington et al. 1999) which shows some
fraction to have very red colors consistent with them
being high redshift galaxies ($z >$ 2). Figure 5 shows
two such optically unidentified radio objects.

\begin{figure}
\centerline{
\epsfig{file=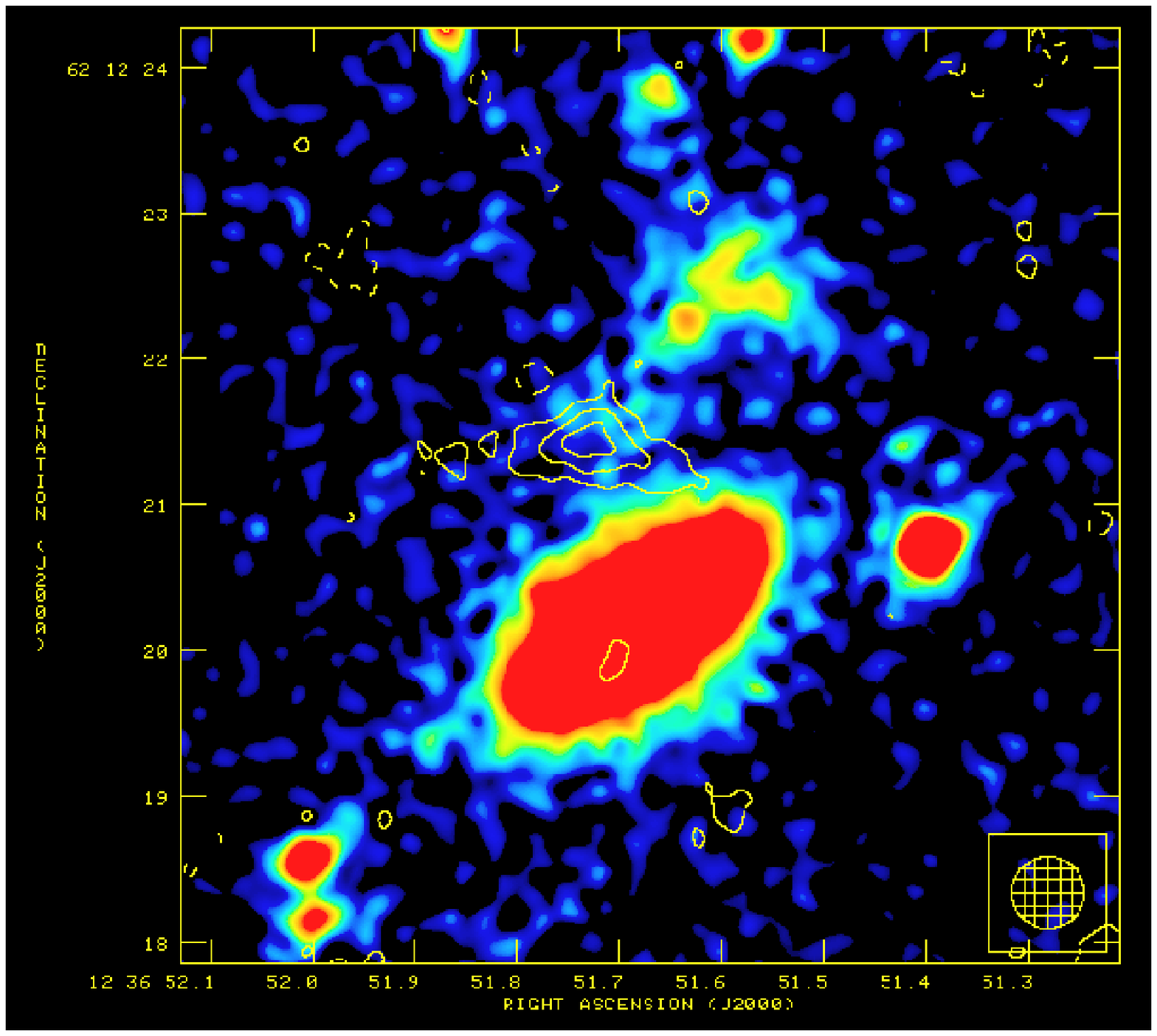,width=20pc}\qquad
\epsfig{file=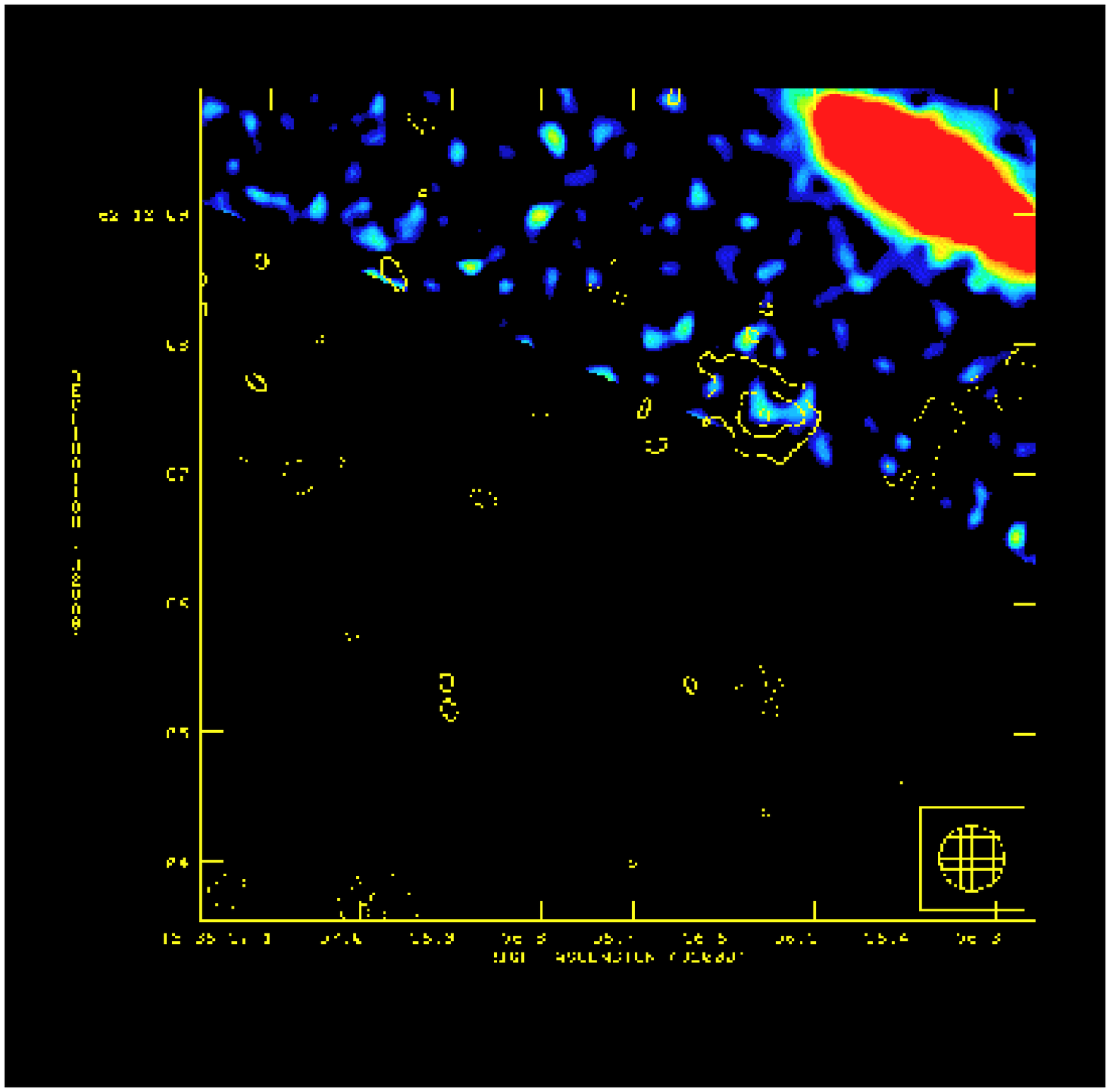,width=15pc}}
\caption{Left) Radio contours for VLA J123651+621221 are
overlaid on the HDF-I image. The sub-mm source
HDF850.1 lies 6\arcs to the northeast.
Right) VLA J123657+621206 is located at the very edge
of the HDF but remains unidentified to I =28.
The source is heavily resolved by the MERLIN 
beam, but is firmly detected in the low resolution
VLA data alone (Richards 1999b). SCUBA source
HDF850.1 is located only 3\arcs away, suggesting
the two are the same object. }

\end{figure}

	Recently, the HDF has been imaged to the confusion
limit by the JCMT/SCUBA at 850 $\mu$m (Hughes et al. 1998).
 Intriguingly, 
two of the radio sources unidentified in the HDF
lie within a few arcsec of the brightest two 
sub-mm sources. This
is especially significant given that the SCUBA 
beam is 15\arcs in diameter. 
VLA J123651+621221 is 
a steep spectrum radio source contained in both
the complete 1.4 and 8.5 GHz samples and is partially
resolved by the 0.2\arcs MERLIN beam. Dickinson
(private communication)
reports a very red object at the position of the
radio source giving further evidence to its 
unusual nature.

	In Cycle 7, we obtained NICMOS imaging of
our brightest unidentified radio object (VLA J123642+621331)
in J and H filters (Waddington et al. 1999).
We obtained a firm detection in H, yielding a color
of I - H $>$ 3.3. Surprisingly, the light profile of
the underlying galaxy is exponential with a half-light
radius of 0.13\arcs . This is strongly suggestive of a
nuclear starburst galaxy at substantial redshift.

	What are these unidentified radio objects?
We consider four possibilities:

1. moderate {\em z} ellipticals - Red ellipticals
   (I - K $>$ 4) at redshifts 1-2 are not
   uncommon. However, the radio
   sources under discussion would have to
   be associated with particularly 
   underluminous or dusty parent ellipticals.
   VLA J123642+621331 clearly does not fall
   into this category.

2. one sided radio jets: We cannot discount the
   possibility that some of our radio sources
   are the brightest jet of a nearby but
   displaced optical galaxy. In this case
   we would expect the parent galaxy to be
   a luminous elliptical with an AGN.
   We find no obvious candidates
   for this scenario.
   
3. very high-$z$ AGN ($z >6$) - Another possibility, is
   that some of the radio sources with very faint 
   optical fluxes are at extreme 
   redshifts, where the Lyman break blanks
   out the I continuum, placing
   them at $z >$ 6. With
   our present radio sensitivity, we could
   have detected a nominal FR I galaxy 
   to approximately $z$ = 10. However, 
   in several cases the radio emission is
   so resolved, it is unlikey to emanate
   from a compact AGN. On the other hand,
   at least two radio objects in the
   HDF remain unidentified to H = 26 
   (Dickinson, private communication)
    and H = 28 (Thompson et al. 1999).

4. high $z$-starbursts (1 $< z <$ 3): Given
   that the majority of submillijansky radio
   sources are associated with star-forming
   galaxies, it is plausible that 
   a tail of the parent galaxy population
    is so obscured by dust that only
   the radio emission is visible. In
   this case we would expect the radio
   emission to be steep spectrum (which
   it is in 18/19 of our objects) and
   the underlying galaxy to be a very red
   disk galaxy. Sevral of our objects
   best fit this description and we consider
   it the most plausible physical explanation. 

	We note that the surface density of these 
objects is about 0.1 square arcmin. There is 
likely some overlap with the faint sub-millimeter
population, although at this point the numbers
are too sparse to make any definitive statements.
Further observations at near infrared and sub-millimeter
wavelengths are necessary to discern the nature
of the optically unidentified radio population.

\section{Conclusions and Future Directions}

	We have shown that the microjansky radio
population is associated primarily with 
star-forming galaxies. The clues that
point to the radio emission being
related to star-formation are 1) the
steep radio spectra, 2) the small, but 
extended angular sizes of the sources, and
3) the identification with luminous disk
galaxies. We have also 
isolated a population
of optically faint radio sources (I $>26 - 28$)
which are possibly distant protogalaxies.

\begin{figure}
\centerline{
\epsfig{file=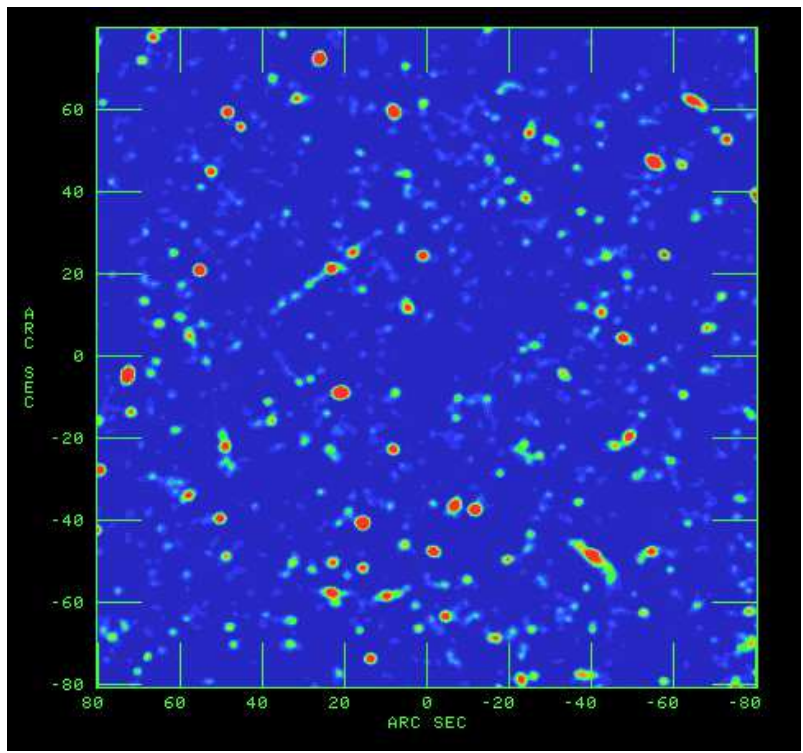,width=25pc}
\caption{Shown is a simulation of a
confusion limited HDF sized region at 1.4 GHz
as seen by the proposed expanded VLA.
The true field of view is 200 times larger,
with a total of 40,000 detections above
0.5 $\mu$Jy (a factor of 80 deeper than the
current VLA/HDF survey). Such observations
could detect the Milky Way at $z = 1$ and
Arp 220 up to $z = 10$ free from dust
obscuration.}}

\end{figure}

        Radio observations provide a powerful
tool in the study of star-formation to the
earliest epochs. 
Together with deep near infrared
and sub-mm observations, they have
the potential to uncover all star-forming galaxies
out to $z \sim$ 2, free from dust extinction.
The  radio luminosity function of star-forming
galaxies at moderate to high$-z$ may ultimately
reveal the  global star-formation history,
free from optical selection biases.
The next generation of centimeter
wavelength telescopes (the expanded VLA, Square Kilometer
Array) will extend our knowledge of the radio properties
of distant galaxies to redshifts of
about 10. Figure 7 shows what the radio sky may
look like at the nanojansky level.

	I wish to thank the LOC for their generous
financial support during the meeting.
I thank my collaborators T. Muxlow, K. Kellermann,
E. Fomalont, B. Partridge, R. Windhorst and I. 
Waddington.

\end{article} 

\end{document}